# Graphene/Ni(111) System: Spin- and Angle-Resolved Photoemission Studies

Yu. S. Dedkov[1], M. Fonin[2], U. Rüdiger[2], and C. Laubschat[1]

[1]Institut für Festkörperphysik, Technische Universität Dresden, 01062 Dresden, Germany
[2]Fachbereich Physik, Universität Konstanz, 78457 Konstanz, Germany

**Here we report spin- and angle-resolved photoemission studies of the inert graphene layer formed on the surface of a ferromagnetic material, Ni(111). X-ray photoemission and spin-resolved spectroscopy of secondary electrons reveal that graphene behaves like a protection layer on the Ni(111) surface preventing its reaction with adsorbed oxygen. Angle-resolved photoelectron spectroscopy of $\pi$ states of graphene on Ni(111) shows a strong dependence of binding energy of these states on the direction of magnetization of the sample. We conclude that the observed extraordinary high "splitting" up to 225 meV of the $\pi$ band in the graphene layer is a manifestation of the Rashba effect which provides a direct possibility to a flexible control of an electron spin in a graphene-based spin-FET.**

THE ability to manipulate a single-electron spin using external electric field is one of the long-sought goals in attempt to develop alternative device technologies. The control over electron spins would be invaluable for nanoscopic spintronics [1]-[3] and related applications, such as quantum information processing. The main paradigm for researches in this field is a spin-based field effect transistor (spin-FET) device introduced by Datta and Das [4] where the spin of injected electron can be modulated via application of only electrical field without any additional magnetic field. The background concept of this effect was realized early on by Rashba [5] showing that an electric field acts as an additional magnetic field in the rest frame of a moving electron causing an energy splitting or spins precession. An ideal spin-FET should combine an effective spin transfer with spin manipulation via electrical field.

However, a high efficient spin-FET device requires long spin relaxation times compared to the mean time of transport through the channel combined with a sufficient difference of the spin rotation angles between two states ("0" and "1") as well as an insensitivity of spin rotation to the carrier energy. In the recent theoretical work [6] spin-FET was proposed on the basis of a single graphene layer, a novel material consisting of a flat monolayer of carbon atoms packed in a two-dimensional honeycomb-lattice, in which the electron dynamics is governed by the Dirac equation [7], [8]. Long electronic mean paths [8] and negligible spin-orbit coupling of the carbon based system [9], i.e. large spin relaxation times, make graphene a best-choice material for the observation of near ballistic spin transport. Here we report on spin- and angle-resolved photoemission studies of high quality graphene layer on the surface of ferromagnetic material, Ni(111).

Experiments were performed in two separated experimental stations in equal conditions. For the present studies graphene layer was prepared via cracking of propene gas ($C_3H_6$) on the Ni(111) surface according to the recipe described elsewhere [10], [11]. A well ordered Ni(111) surface was prepared by thermal deposition of Ni films with a thickness of about 200 Å onto a clean W(110) substrate and subsequent annealing at 300°C. Spin- and angle-resolved photoemission spectra were recorded at 1486.6 eV (XPS) and 40.8 eV (UPS) photon energies, respectively, at room temperature using a hemispherical energy analyzer SPECS PHOIBOS 150 combined with a 25 kV mini-Mott spin-detector. The energy resolution of the analyzer was set to 50 and 500 meV for UPS and XPS, respectively. The spin polarization of secondary electrons in XPS spectra was analyzed with an energy resolution of 200 meV. Spin- and angle-resolved photoemission measurements were performed in remanence after having applied a magnetic field pulse of about 1 kOe along the in-plane <1-10> easy axis of the Ni(111) film. The experimental setup asymmetry in spin polarization analysis was accounted for by measuring spin-resolved spectra for two opposite directions of applied field [12]. Scanning tunneling microscopy (STM) measurements were performed in ultrahigh vacuum with an Omicron VT AFM/STM at room temperature using electrochemically etched tungsten tips that were flash annealed by electron bombardment.

STM images of high-quality graphene layer formed on the Ni(111) surface are shown in Fig. 1 (a,b). After the cracking procedure the Ni(111) surface is completely covered by a graphene film. All investigated terraces display the same atomic structure as discussed in details in [13], [14]. Extracted value of the distance between carbon atoms in graphene layer is 2.4±0.1 Å which is in good agreement with the expected interatomic spacings of 2.46 Å in graphene. The LEED image of the graphene/Ni(111) system (inset in Fig. 1(a)) reveals a well-ordered $p(1\times1)$-overstructure as expected from the small lattice mismatch of only 1.3%.

The electronic structure of the graphene/Ni(111) was studied in detail by means of angle-resolved photoemission in earlier work [11], [13]-[15]. Here, we show angle-resolved photoemission spectra of the system under study taken with 40.8 eV along the $\Gamma-M$ direction of the surface Brillouin zone (Fig. 1 (c)). For two dimensional systems, like graphene, the energy dispersion can be described as a function of only the in-plane component of the wave vector, $k_\parallel$. In the following, describing $\pi$ states of graphene we will not distinguish between $k$ and $k_\parallel$. From a comparison of the photoemission spectra of graphene/Ni(111) system with pure graphite [16] we conclude that the difference in the binding energy of the π-states is about 2.3 eV which is close to the value observed earlier [11], [13]-[15] and in good agreement with the theoretical prediction of 2.35 eV [17]. This effect reflects the effect of strong hybridization of the graphene π-states with the Ni 3$d$ bands.

The inert properties of the graphene/Ni(111) system were tested by the exposure to oxygen for 30 min at a partial $O_2$-pressure of $5\times10^{-6}$ mbar and room temperature. The results are shown in Fig. 2: (a) series of Ni 2$p$ XPS spectra taken from pure Ni(111) (spectrum 1), from freshly prepared graphene/Ni(111) (spectrum 2), and after exposure of graphene/Ni(111) and Ni(111) to oxygen (spectra 3 and 4, respectively); (b) spin polarization of the secondary electrons measured for the aforementioned systems (as marked in the Figure). The inset in Fig. 2 (a) shows O 1$s$ XPS spectra obtained from the latter two systems. In all spectra shown in Fig. 2 (a), the Ni 2$p$ emission consists of a spin-doublet (2$p_{3/2,1/2}$) and a well-known satellite structure. For the pure Ni(111) film, the satellite appears with respect to the main lines at 6 eV higher binding energy, whereas this shift is increased approximately by 0.9 eV for the graphene/Ni(111) system. This effect reflects the altered chemical environment at the interface and is not the subject of the present discussion. From the comparison of spectra 2 and 3 in Fig. 2 (a), it becomes clear that the exposure to oxygen does not affect the spectra shape of the graphene/Ni(111) system. It shows that the graphene overlayer prevents obviously the interaction of oxygen with the underlying Ni substrate which would be reflected by strong modification of satellite structure of the Ni 2$p$ spectra (spectrum 4). From the observation of weak O 1$s$ photoemission signal of the graphene/Ni(111) system after the oxygen exposure we conclude that the sticking coefficient of oxygen on the graphene overlayer is extremely low and the overlayer is almost free of defects that may allow oxygen atoms to access the Ni substrate.

The spin polarization of secondary electrons emitted in XPS is shown in Fig. 2 (b) for all systems discussed above. The spin polarization of the low kinetic energy tail of the photoelectron spectrum of the ferromagnetic Ni(111) film is in good agreement with previous results [18], [19]. The spin polarization at kinetic energies of 10 eV and above is equal to the one of the valence band. A strong enhancement of spin polarization appearing at 3 eV is due to the difference in the inelastic mean free path for spin-up and spin-down electrons which scatter to the spin-polarized empty states of the ferromagnetic material in the close vicinity of the Fermi level. The spin polarization of secondary electrons at zero kinetic energy for the graphene/Ni(111) system is about 12% compare to 17% for the clean Ni(111) surface. This reduction is due to the strong hybridization between Ni 3$d$ and graphene π states [17]. The spin polarization of secondary electrons can be taken as proportional to the magnetic moment of Ni atoms at the surface [18], [19]. In this case one can estimate that in the graphene/Ni(111) system the magnetic moment of Ni atoms at the interface amounts to about $0.52\mu_B$ as compared to $0.72\mu_B$ for the pure Ni(111) surface [17]. The exposure of graphene/Ni(111) to oxygen at the conditions mentioned above does not influence on the shape of the spin polarization curve. Practically, the spin polarization at zero kinetic energy remains almost the same. This is in strong contrast to the behavior of the clean Ni surface where exposure to oxygen leads to a complete vanishing of the spin polarization (see Fig. 2 (b)).

The studies of the Rashba effect were performed in the geometry discussed elsewhere [14]: angle-resolved photoemission spectra are measured for two opposite magnetization directions of the sample. Hybridization between Ni 3$d$ and graphene π states leads to the strong effective electric field (**E**) at the graphene/Ni interface. The experimental geometry is chosen in the way that three vectors (**E**, **k**, **s**) are orthogonal each other (**k** and **s** are the wave vector and the spin of an electron) allowing for the observation of the Rashba effect in the studied system. Rashba Hamiltonian can be written as $H_R=\alpha_R(\mathbf{E}\times\mathbf{k})\cdot\mathbf{s}$ which leads to the different binding energy for electrons with different **s** and corresponding splitting of energy bands is proportional to the electron wave vector $k$: $\Delta E(\mathbf{k})=E(\mathbf{k},+\mathbf{M})-E(\mathbf{k},-\mathbf{M})$. Here, $\pm\mathbf{M}$ denotes two opposite directions of magnetization which is connected with spin of electron, **s**.

In Fig. 3 (a) we show two representative pairs of photoemission spectra measured for two opposite magnetization directions and for the emission angles θ=0° and θ=24° corresponding to the wave vectors $k=0$ Å$^{-1}$ and $k=1.16$ Å$^{-1}$, respectively. A clear shift of $\Delta E\approx200$ meV in the peak position of the π states upon magnetization reversal is observed for spectra taken in off-normal emission geometry, whereas no shift is observed for spectra measured at normal emission. Although the states for opposite magnetization directions do not exist simultaneously (Fig. 3 (a)), the observed energy shift is equivalent to the Rashba splitting at a nonmagnetic surface, and we shall refer to it further as the Rashba "splitting". In Fig. 3 (b) we plot the magnitude of the Rashba "splitting" $\Delta E(\mathbf{k})$ extracted from the energy dispersion of graphene π states as a function of the wave vector $k$ along the Γ−M direction of the surface Brillouin zone. The linear dependence of $\Delta E$ on $k$ is a clear proof of the Rashba effect in a graphene layer on Ni(111). The Rashba "splitting" disappears on approaching the M point at the border of the surface Brillouin zone. This can be explained by the fact that the branches have to exchange their energy positions by crossing the border of the Brillouin zone. The same effect is expected in all other directions. The experimental proof for the other high-symmetry direction, Γ−K, is not possible in this experimental geometry since the sample is magnetized along the <1-10> easy axis of magnetization of the Ni(111) substrate which is parallel to Γ−K. For practical use of the effect all energy branches should be pushed to lower binding energies via hole-doping (for example, iodine or $FeCl_3$ [20]) with a sizable potential gradient in order to produce the detectable Rashba "splitting". The weak magnetic linear dichroism effects visible at $\theta=24°$ for the Ni 3*d* states leads only to changes in photoemission intensities and can not be responsible for the energy shift of the band. Moreover, such dichroism effects can not be detected for the π states of graphene because of an extremely weak spin-orbit interaction in carbon.

The large Rashba "splitting" of the π states observed in this experiment is thus with high probability due to the asymmetry of the charge distribution at the graphene/Ni(111) interface. The hybridization between the π states of graphene and the Ni 3*d* states leads to a considerable charge transfer from the 3*d* states of the Ni surface atoms to the unoccupied $π^*$ states of the graphene monolayer [21]. As a consequence, a large gradient of potential is formed at the interface and the electrons in the interface state are subject to an effective spatially averaged crystal electric field that leads to the Rashba effect. A similar effect was recently observed at the oxygen/Gd(0001) interface [22], where *ab initio* calculations excluding the spin-orbit interaction demonstrate that in this case the gradient of the potential plays the crucial role as the origin of the Rashba "splitting". However, this model can be implemented for the graphene/Ni(111) system only in the case of the spin-polarized π band of graphene. The spin-polarized carriers in the graphene layer can be due to hybridization or the proximity effect [23], [24]. Since the Fermi surface of graphene is centered around the K point in reciprocal space where the Ni substrate has states of pure minority spin character [23] one may expect that predominantly spin-down electrons will fill $π^*$ states and produce the observed spin polarization of these states. We conclude that the observed Rashba effect in this system is due to the interaction of spin-polarized electrons in the π band of the graphene layer with the large effective electric field at the graphene/Ni(111) interface.

In conclusion, the electronic structure and magnetic properties of high quality graphene layer prepared via cracking of propene gas on Ni(111) surface was studied by means of angle-resolved photoemission and spin-resolved spectroscopy of true secondary electrons. It was fond that graphene layer protects underlying Ni surface against oxidation and we conclude that such inert graphene/Ni(111) system can be used as a stable source of spin-polarized electrons. Angle-resolved photoemission studies demonstrate the difference in the binding energy of the graphene π states depending on the magnetization direction of the Ni film. These observations are explained as a manifestation of the Rashba effect in graphene/Ni(111) system. Our findings show that an electron spin in the graphene layer can be manipulated in a controlled way and have important implications for graphene-based spintronic devices.

ACKNOWLEDGMENT

This work was funded by the Deutsche Forschungsgemeinschaft (DFG) through Collaborative Research Centers SFB-463 (Projects B4 and B16) and SFB-767 (Project TP C5).

REFERENCES

[1] S. A. Wolf *et al.*, "Spintronics: A spin-based electronics vision for the future," *Science*, vol. 294, pp. 1488-1495, 2001.
[2] *Spin Electronics*, edited by D. Awshalom, Dordrecht: Kluwer, 2004.
[3] S. A. Wolf, A. Y. Chtchelkanova, and D. M. Treger, "Spintronics-A retrospective and perspective," *IBM J. Res. Dev.*, vol. 50, pp. 101-110, 2006.
[4] S. Datta and B. Das, "Electronic analog of the electro-optic modulator," *Appl. Phys. Lett.* vol. 56, pp. 665-667, 1990.
[5] Yu. A. Bychkov and E. I. Rashba, "Oscillatory effects and the magnetic susceptibility of carriers in inversion layers," *J. Phys. C: Solid State Phys.*, vol. 17, pp. 6039-6045, 1984.
[6] Y. G. Semenov, K. W. Kim, and J. M. Zavada, "Spin field effect transistor with a graphene channel," *Appl. Phys. Lett.*, vol. 91, pp. 153105, 2007.
[7] M. Wilson, "Electrons in atomically thin carbon sheets behave like massless particles," *Physics Today*, vol. 59, pp.21-23, 2006.
[8] A. K. Geim and K. S. Novoselov, "The rise of graphene," *Nature Materials*, vol. 6, pp. 183-191, 2007.
[9] H. Min, J. E. Hill, N. A. Sinitsyn, B. R. Sahu, L. Kleinman, and A. H. MacDonald, "Intrinsic and Rashba spin-orbit interactions in graphene sheets," *Phys. Rev. B*, vol. 74, pp. 165310, 2006.
[10] D. Farias, A. M. Shikin, K.-H. Rieder, and Yu. S. Dedkov, "Synthesis of a weakly bonded graphite monolayer on Ni(111) by intercalation of silver," *J. Phys.: Condens. Matter*, vol. 11, pp. 8453-8458, 1999.
[11] Yu. S. Dedkov, A. M. Shikin, V. K. Adamchuk, S. L. Molodtsov, C. Laubschat, A. Bauer, and G. Kaindl, "Intercalation of copper underneath a monolayer of graphite on Ni(111)," *Phys. Rev. B*, vol. 64, pp. 035405, 2001.
[12] J. Kessler, *Polarized Electrons*, 2nd ed. Berlin: Springer, 1985.
[13] Yu. S. Dedkov, M. Fonin, and C. Laubschat, "A possible source of spin-polarized electrons: The inert graphene/Ni(111) system," *Appl. Phys. Lett.*, vol. 92, pp. 052506, 2008.
[14] Yu. S. Dedkov, M. Fonin. U. Rüdiger, and C. Laubschat, "Rashba effect in the graphene/Ni(111) system," *Phys. Rev. Lett.*, to be published.
[15] A. Nagashima, N. Tejima, and C. Oshima, "Electronic states of the pristine and alkali-metal-intercalated monolayer graphite/Ni(111) systems," *Phys. Rev. B*, vol. 50, 17487-17495 (1994).
[16] A. M. Shikin, S. L. Molodtsov, A. G. Vyatkin, V. K. Adamchuk, N. Franco, M. Martin, and M. C. Asensio, "Electronic structure of surface compounds formed under thermal annealing of the La/graphite interface," *Surf. Sci.*, vol. 429, pp. 287-297, 1999.
[17] G. Bertoni, L. Calmels, A. Altibelli, and V. Serin, "First-principles calculation of the electronic structure and EELS spectra at the graphene/Ni(111) interface," *Phys. Rev. B*, vol. 71, pp. 075402, 2005.
[18] H. Hopster, R. Raue, E. Kisker, G. Güntherodt, and M. Campagna, "Evidence for Spin-Dependent Electron-Hole-Pair Excitations in Spin-Polarized Secondary-Electron Emission from Ni(110)," *Phys. Rev. Lett.*, vol. 50, pp. 70-73, 1983.
[19] R. Allenspach, "Spin-polarized scanning electron microscopy," *IBM J. Res. Dev.*, vol. 44, pp. 553-570, 2000.
[20] M. S. Dresselhaus and G. Dresselhaus, "Intercalation compounds of graphite," *Adv. Phys.*, vol. 30, pp. 139-326, 1981.
[21] K. Yamamoto, M. Fukushima, T. Osaka, and C. Oshima, "Charge-transfer mechanism for the (monolayer graphite)/Ni(111) system," *Phys. Rev. B*, vol. 45, pp. 11358-11361, 1992.
[22] O. Krupin, G. Bihlmayer, K. Starke, S. Gorovikov, J. E. Prieto, K. Döbrich, S. Blügel, and G. Kaindl, "Rashba effect at magnetic metal surfaces," *Phys. Rev. B*, vol. 71, pp. 201403, 2005.
[23] V. M. Karpan, G. Giovannetti, P. A. Khomyakov, M. Talanana, A. A. Starikov, M. Zwierzycki, J. van den Brink, G. Brocks, and P. J. Kelly, "Graphite and graphene as perfect spin filters," *Phys. Rev. Lett.*, vol. 99, pp. 176602, 2007.
[24] P. M. Tedrow, J. E. Tkaczyk, and A. Kumar, "Spin-polarized electron tunneling study of an artificially layered superconductor with internal magnetic field: EuO-Al," *Phys. Rev. Lett.*, vol. 56, pp. 1746-1749, 1986.

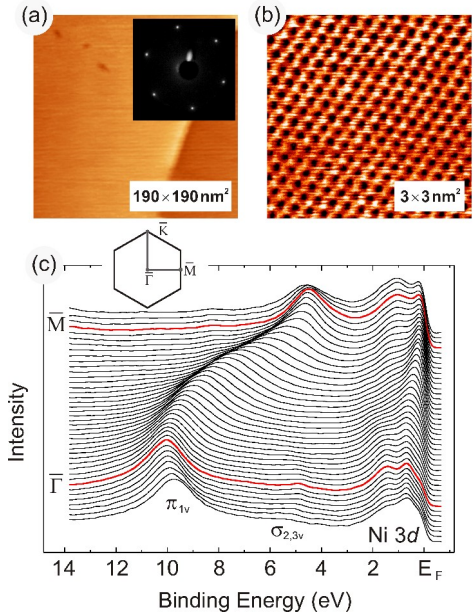

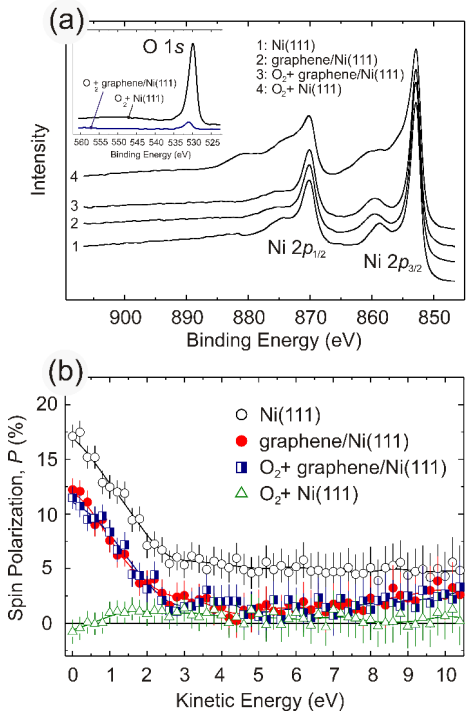

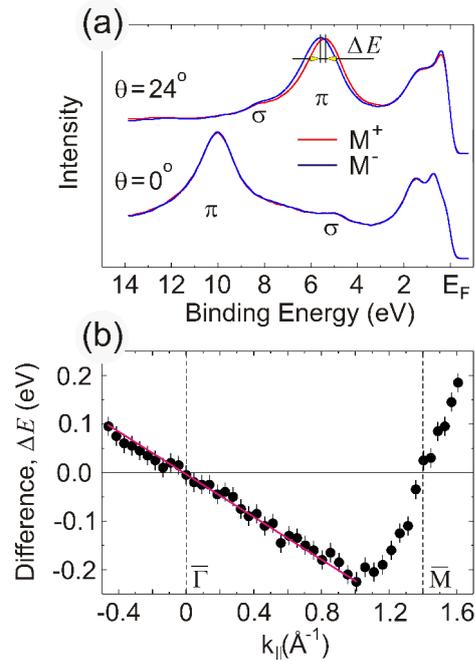

Fig. 1. (a,b) Constant current STM imagea of the graphene/Ni(111) surface. The inset of (a) shows a LEED image obtained at 63 eV. (c) Angle-resolved photoemission spectra measured with hν=40.8 eV along Γ−M direction of the surface Brillouin zone.

Fig. 2. (a) Ni 2$p$ core-level spectra of Ni(111), graphene/Ni(111), and systems obtained after exposure of the respective surfaces to oxygen. The inset shows O 1$s$ XPS spectra obtained after exposure of Ni(111) and graphene/Ni(111) surfaces to large amounts of oxygen, respectively. (b) Spin-polarization of secondary electrons from Ni(111), graphene/Ni(111), O$_2$+graphene/Ni(111), and O$_2$+Ni(111) (open triangles) systems after excitation with Al Kα radiation. Solid lines are shown as guides to the eye.

Fig. 3. (a) Series of representative photoelectron spectra of the graphene/Ni(111) system for two different emission angles (marked in the plot) and two directions of magnetization, respectively. (b) Rashba "splitting" $\Delta E(\mathbf{k})$ obtained from the angle-resolved photoemission data as a function of the wave vector $k$ along the Γ−M direction of the surface Brillouin zone.